\documentclass[aps,12pts,prl,showpacs,amsmath,amssymb,preprint]{revtex4}
\usepackage{graphicx}
\usepackage{bm}
\usepackage{subfigure}
\usepackage{mathptmx}
\usepackage{epstopdf}
\usepackage[usenames,dvipsnames]{color}
\begin{document}
\title{Novel magnetic arrangement and structural phase transition\\
 induced by spin-lattice coupling in multiferroics}
\author{Satadeep Bhattacharjee$^{1}$, Dovran Rahmedov$^{1}$, Dawei Wang$^{2}$ and Laurent Bellaiche$^{1}$}
%\email{sbhattac@uark.edu}
\affiliation{$^{1}$Department of Physics and Institute for Nanoscience and Engineering, 
University of Arkansas, Fayetteville, Arkansas 72701, USA \\
$^{2}$Electronic Materials Research Laboratory, Key Laboratory of the
  Ministry of Education and International Center for Dielectric
  Research, Xi'an Jiaotong University, Xi'an 710049, China}
  
  \begin{abstract}
Using an  effective Hamiltonian of mutiferroic BiFeO$_3$ (BFO) as a toy model, we explore the
effect of the  coefficient, $C$, characterizing the strength of the spin-current interaction, on physical properties.
We observe that for larger $C$ values and below a critical temperature, the magnetic moments organize themselves in a novel cycloid which propagates along a low-symmetry direction and is associated with a structural phase transition from polar rhombohedral to a polar triclinic state. We emphasize that both of these magnetic and structural transitions are results of a remarkable self-organization of {\it different} solutions of the spin-current model.
 \end{abstract}
\maketitle

Multiferroic materials form an exciting type of materials which can have multiple ferroic properties in the same phase \cite{MF1,MF2} such as (anti)ferromagnetism, ferroelectricity, ferroelasticity etc...
Out of the different kinds of multiferroics,  the ones possessing both ferroelectric and magnetic orders have drawn particular attention \cite{J1,J2,C1,C2}. This is because the  magneto-electric coupling between these two orders can lead to the control of magnetism by electric field, or, vice-versa, to the manipulation of electric properties by the application of magnetic field (such control is attractive for the design of original devices
and  is also of academic interest).
One particularly known and important example of such coupling is provided by the
so-called spin-current model \cite{Katsura1,Aldo} for which the interaction energy is given by:
\begin{align}
\Delta E = -C ( {\bf P} \times {\bf e_{ij}}) \cdot (\bf{m_i}\times \bf{m_j})~~~~,
\end{align}
where $P$ is  the electric polarization, ${\bf e}_{ij}$ is the unit vector along a specific direction joining site $i$ to site $j$ and where $ {\bf m}_i$ and
 $\bf{m_j}$ are the magnetic moments located at these sites $i$ and $j$, respectively. $C$ is a coefficient characterizing the strength of this spin-current interaction \cite{DM}. For instance, the spin-current model was advocated \citep{Katsura1} to explain the occurrence of a spontaneous electric polarization at the temperature at which Mn spins form some spiral order in orthorhombically distorted multiferroics such as TbMnO$_3$, DyMnO$_3$, and GdMnO$_3$  \cite{RMnO3}. Similarly, the spin-current model has been proposed as one possible mechanism behind  the clear
correlation between the spin-helicity and the electronic polarization found in neutron diffraction measurements in Gd$_{0.7}$Tb$_{0.3}$MnO$_3$ \cite{helical}. 
This spin-current model has also been recently shown \cite{PRL1,excite1} to be responsible for the polarization-induced formation of the magnetic cycloid \cite{Scott,excite1,Sosnowska1,Sosnowska2,Zalesskii2002,Zalesskii2000,Bush,RamazanogluPRB2011,Sosnowska2011,Sosnowska2006,Pokalitov,Zvedin,deSouza,Jeong}.
 It is of common wisdom to consider that (long-time-sought and desired)  high magneto-electric coupling will be achieved when finding systems possessing a large value for the $C$ coefficient appearing in Eq. (1). However, in that situation, it is also legitimate to ask two other important and currently unresolved questions, that are:
 (1) can large $C$ values give rise to novel types of magnetic organization with respect to those of small or intermediate $C$? and (2) is the structural ground state affected by a large $C$ coefficient?
 If yes, determining the microscopic origins of such effects is also of obvious importance.
     
The aim of this Letter is to address these issues, by varying the strength of the spin-current
interaction in the effective Hamiltonian approach of BFO (which is our present toy model).
 As we will see, surprises are in store.
For instance, large $C$ values lead to the formation of a previously unknown magnetic cycloid that propagates along an unusual low-symmetry direction. They also generate a structural phase transition from a polar high symmetry phase to a polar low-symmetry state, that is associated with a magnetically-induced rotation of the electrical polarization. These striking features arise from a remarkable self-organization between {\it different} individual solutions of the spin-current model.

Let us first recall that the total energy corresponding to the effective Hamiltonian of BFO is expressed as \cite{PRL1}: 
\begin{align}
E_{tot}= E_{FE-AFD}\biggl(\{{\bf u}_i\},\{{\bf \eta}_i\},\{{\bf \omega}_i\}\biggr)\nonumber+\nonumber\\
E_{Mag}\biggl(\{{\bf m}_i\},\{{\bf u}_i\},\{{\bf \eta}_i\},\{{\bf \omega}_i\}\biggr)~~~~,
\end{align}
where the first part contains all non-magnetic variables: the local mode (${\bf u}_i$), that is directly proportional to the electric dipole centered in the unit cell $i$ \cite{Zhong}; the strain in this unit cell
(${\bf \eta}_i$) that gathers both homogeneous and inhomogeneous contributions \cite{Zhong}; and the antiferrodistortive rotational mode (${\bf \omega}_i$) that is associated with oxygen octahedral tilting in unit cell i. All these variables are centered on  Fe sites.
The second term in the total 
energy is the magnetic-dependent
term. It includes the mutual interaction between magnetic moments of Fe ions at different cells $i$ (${\bf m}_i$),  that all have a fixed magnitude of 4$\mu_B$. It also contains  interactions between magnetic moments and the other degrees of freedom (namely, local modes,  antiferrodistortive motions and strains). The analytical form of $E_{Mag}$ is same as is Ref.[\onlinecite{PRL1}], and is thus given by:
\begin{align}
E_{Mag}\biggl(\{{\bf m}_i\},\{{\bf u}_i\},\{{\bf \eta}_i\},\{{\bf \omega}_i\}\biggr)= \sum_{i,j,\alpha,\gamma}Q_{ij,\alpha\gamma}{\bf m}_{i,\alpha}
{\bf m}_{j,\gamma}\nonumber\\+\sum_{i,j,\alpha,\gamma}D_{ij,\alpha\gamma}{\bf m}_{i,\alpha}{\bf m}_{j,\gamma}+
\sum_{i,j,\alpha,\gamma}E_{ij,\alpha\gamma,\nu\delta}{\bf m}_{i,\alpha}{\bf m}_{j,\gamma}u_{i,\nu}u_{i,\delta}\nonumber\\+
\sum_{i,j,\alpha,\gamma}F_{ij,\alpha\gamma,\nu\delta}{\bf m}_{i,\alpha}{\bf m}_{j,\gamma}\omega_{i,\nu}\omega_{i,\delta}+
\sum_{ijl\alpha\gamma}G_{ijl\alpha\gamma}\eta_{l}(i){\bf m}_{i,\alpha}{\bf m}_{j,\gamma}
\nonumber\\
\sum_{i,j} K_{ij}(\omega_i-\omega_j)\cdot({\bf m}_i\times {\bf m}_j)-C \sum_{ij}({\bf u}_i\times {\bf e_{ij}})\cdot({\bf m}_i\times {\bf m}_j)
\end{align}
The first term in the above expression represents magnetic dipolar interaction where both indices i and j run over all the
sites. The second term is the direct magnetic exchange between magnetic moments at sites i and j. The third, fourth and fifth
terms characterize the change in magnetic exchange interaction induced by the local modes,  antiferrodistortive motions and strains. The index j
for the second, third, fourth and fifth term runs over first, second and third nearest neighbors of site $i$. 
The first five energies desire to induce a collinear magnetism. On the other hand, the sixth term, in which the j index runs over the six first nearest neighbors of site $i$, is at the origin of  the spin-canted weak ferromagnetic structure of BFO, and involve the  rotations of the oxygen octahedra \cite{Al,ca,Zhigang,Daweidyn}. The last term represents the spin-current model \cite{Katsura1,Aldo}, and has been found to be essential to  reproduce the complex cycloidal structure of BFO bulks \cite{PRL1}.  In order to appreciate the results of this present study (to be discussed below), it is important to know that, for this last term, $j$ runs over the twelve {\it second nearest neighbors} of site $i$, and {\bf e}$_{ij}$ is the unit  vector along the direction joining site $i$ to site $j$ (for symmetry reasons, this sum has to run over {\it all} second-nearest-neighbor directions, since the reference state of our Taylor expansion of $E_{tot}$ is  the cubic paraelectric and paramagnetic state). Here, the
coefficient C appearing in that last term is allowed to vary, in order to find the effect of such spin-lattice coupling on physical properties. As we will see, this variation can lead to new magnetic configurations, as well as anomalous behavior of the electrical polarization.
\par
We have performed Monte Carlo (MC) simulations using the $E_{tot}$  internal energy with a $12\times 12\times 12$ supercell.
The finite-temperature properties of the modeled bulk perovskite system  are obtained using $10^6$ MC sweeps for equilibration
and $10^5$ additional sweeps  to calculate statistical averages.

Figure 1 schematizes the different magnetic structures we numerically found at low temperature, when varying the $C$ parameter. 
Below the critical value of $1.2\times 10^{-5}$ Hartree/Bohr$\mu_B^2$, the magnetic ground state is the spin-canted structure reported in Refs. \cite{Al,ca,Zhigang}, for which a large G-type antiferromagnetic vector coexists with a weak and perpendicular ferromagnetic vector. 

Above this value and up to a coefficient $C$ $\sim 4.2\times 10^{-5}$ Hartree/Bohr$\mu_B^2$, the resulting magnetic dipoles organize themselves into the cycloidal configuration that is known to occur in BFO bulk \cite{Scott,excite1,Sosnowska1,Sosnowska2,Zalesskii2002,Zalesskii2000,Bush,RamazanogluPRB2011,Sosnowska2011,Sosnowska2006,Pokalitov,Zvedin,deSouza,Jeong}, that is the magnetic dipoles (mostly) rotate in an $\{\bar{2}11\}$ plane possessing both the electrical polarization (that is oriented along the pseudo-cubic [111] direction) and one direction  (i) joining second nearest
neighbors and (ii) being  {\it perpendicular} to the electrical polarization (such as the pseudo-cubic [0$\bar{1}$1] direction). This direction joining second-nearest
neighbors  is spanned by {\it one} of the twelve {\bf e}$_{ij}$  vectors appearing in the last term of Eq. (3), and coincides with the propagation direction of the 
cycloid \cite{footnoteSDW}. This cycloid will be denoted as ``type-I cycloid'' in the following. 

Surprisingly, for larger values of the $C$ coefficient, a novel magnetic structure emerges. It consists in another cycloidal configuration, to be referred to as ``type-II cycloid''. Its propagation direction is now oriented along an unusual (low-symmetry) 
pseudo-cubic $<0\bar{1}2>$ direction  \cite{FFT}. As seemingly at odds with any of the 12 different sums over $j$ running in the last term of Eq. (3),  
this direction is therefore {\it not} lying anymore along any second-nearest neighbor direction. 
It is also not perpendicular to the polarization (which is lying close to [111])! Furthermore,  
the magnetic dipoles in the type-II cycloid are found to rotate within a $\{\bar{3}21\}$ plane that contains both the polarization and this novel cycloidal propagation direction.

%-----------------------Figure here--------------------------------------------------------------------------------------------------------
%\begin{figure}
%\includegraphics[width=75mm,height=85mm]{../Figs/Fig1.eps}
%\caption{(Color online) Range of the different magnetic structures numerically found when varying the spin-lattice coupling constant C.}
%\label{Fig.1}
%\end{figure}
%------------------------Figure here----------------------------------------------------------------------------------------------------------

Before trying to understand the origin(s) of this original magnetic cycloid, let us also investigate if the electrical 
polarization can be affected by  the $C$ coefficient. Consequently,  Figs. 2(a), (b) and (c) show the evolution 
of the $<{\bf u}>$ supercell average of the local modes as a function of temperature for three different representatives values of $C$. They are, respectively, 
$7\times 10^{-6}$ Hartree/Bohr$\mu_B^2$ (which leads to the spin-canted structure at low temperature),  
$2.8\times10^{-5}$Hartree/Bohr$\mu_B^2$ (which results in the type-I cycloid at low temperature), and  $6.3\times10^{-5}$ 
Hartree/Bohr$\mu_B^2$ (which generates a type-II cycloid at low temperature).
One can see that for the two smallest $C$ coefficients, a paraelectric--to-ferroelectric transition occurs at T$_C$ about 1100K, 
below which the polarization points along the pseudo-cubic [111] direction (since the x-, y- and z-components of $<{\bf u}>$ 
are all equal to each other and non-null). The polarization remains oriented along 
[111] below the magnetic transition temperature, T$_N$ $\simeq$ 660K.
On the other hand, the third situation (corresponding to $C=6.3\times10^{-5}$) is notably different from the first two cases for several reasons. First of all, $T_C$ and $T_N$ significantly increase 
by 96 K and 340 K, respectively. 
Such enhancement can be understood by realizing that the last term of Eq. (3) indicates that increasing the $C$ coefficient favors both the 
formation of local modes (and hence of a polarization) and of the cross product between magnetic moments at sites $i$ and $j$ (and hence of a cycloidal-type structure). 
Secondly, while the polarization remains oriented along the [111] pseudo-cubic direction between $T_N$ and $T_C$, it rotates 
towards a low-symmetry $[pqr]$ direction -- with $r<p<q$ -- below $T_N$ (note that, as shown in Fig. 2c, the x- and y-components of the local modes are close to each other but are physically distinct, 
since their difference is found to be larger than their statistical error bars). In other words, the occurrence of the type-II 
cycloid results in a structural transformation from a polar rhombohedral phase 
to a polar triclinic state! We are not aware that such type of strong magneto-electric effect, 
correlating the formation of magnetic ordering with a lowering of symmetry between polar phases and with rotation of the polarization, 
has ever been previously found (Note that Refs. \cite{excite1,pred2} suggested that the formation of the type-I cycloid in BFO would result in the transformation of a polar rhombohedral to a polar 
monoclinic state, but this interpretation is likely incorrect, since it is now known that BFO bulk remains in a  rhombohedral $R3c$ state  below $T_N$). 
It is also worthwhile to know that the polarization in any of the three cases depicted in Figs. 2 is found to be perfectly homogeneous at low temperature. This implies that the breaking of symmetry between the x-, y- and z-components of the polarization seen in Fig. 2c do {\it not} originate from the formation of domains (having, e.g, small {\it versus} large z-component of the electric dipoles).

\begin{table}
%\begin{tabular}{|c|lccl|}
\begin{tabular}{|c|c|c|c|}
\hline
Direction & Spin-current Energy (Hartree) & Relative Spin-current Energy (dimensionless)  & $<{\bf m}_i\times {\bf m}_j>$ ($\mu^2_B$) \\
\hline
\hline
{[110]} &   -0.26 & 0.05 & -6$\hat{\bf x}$+4$\hat{\bf y}$+2$\hat{\bf z}$\\
{[1$\bar 1$0]} &   -0.19& 0.04 & 6$\hat{\bf x}$-4$\hat{\bf y}$-2$\hat{\bf z}$  \\
{[101]}  &  -0.50  &  0.10 & 12$\hat{\bf x}$-8$\hat{\bf y}$-4$\hat{\bf z}$\\
{[10$\bar 1$]}  &  -0.50 &  0.10 & -12$\hat{\bf x}$+8$\hat{\bf y}$+4$\hat{\bf z}$\\
{[$\bar 1$10]}  &  -0.19  & 0.04 & -6$\hat{\bf x}$+4$\hat{\bf y}$+2$\hat{\bf z}$\\
{[$\bar 1 \bar 1$0]}  &  -0.26 & 0.05 & 6$\hat{\bf x}$-4$\hat{\bf y}$-2$\hat{\bf z}$\\
{[$\bar 1$01]}   & -0.50 &  0.10 & 12$\hat{\bf x}$-8$\hat{\bf y}$-4$\hat{\bf z}$ \\
{[$\bar 1$0$\bar 1$]}  &  -0.50 &  0.10 & -12$\hat{\bf x}$+8$\hat{\bf y}$+4$\hat{\bf z}$  \\
{[011]}   & -0.06 & 0.01 & 6$\hat{\bf x}$-4$\hat{\bf y}$-2$\hat{\bf z}$ \\
{[0$\bar 1$1]} &   -0.96 & 0.20 & 12$\hat{\bf x}$-8$\hat{\bf y}$-4$\hat{\bf z}$ \\
{[01$\bar 1$]}  &  -0.96  & 0.20 &  -12$\hat{\bf x}$+8$\hat{\bf y}$+4$\hat{\bf z}$ \\
{[0$\bar 1\bar 1$]}  &  -0.06  & 0.01 &  -6$\hat{\bf x}$+4$\hat{\bf y}$+2$\hat{\bf z}$ \\
\hline
\end{tabular}
\caption{Low-temperature energies associated with the last term of the effective magnetic Hamiltonian (see text) and averaged ${\bf m}_i\times {\bf m}_j$  for the twelve different second nearest-neighbor directions, when the spin-lattice coupling constant $C=6.3\times10^{-5}$. The Cartesian components of the averaged ${\bf m}_i\times {\bf m}_j$ have been rounded to  integers.
The quantities shown in Table I are those associated with a single snapshot of the type-II cycloid structure in a $12\times 12\times 12$  simulation box.}
\end{table}

Let us now reveal why a type-II cycloid and a resulting low-symmetry structural ground state can occur for large $C$ coefficient. For that, Table I reports, at low-temperature and for  $C=6.3\times10^{-5}$ (i.e., when a type-II cycloid is in-place): (i) the ``spin-current'' energy associated with each of the twelve second-nearest neighbor directions, that is the energy of each of these 12 
directions appearing in the last term of Eq. (3); (ii)  the resulting {\it  relative} ``spin-current'' energy, which is defined as the ``spin-current'' energy divided by the total energy of  the last  term of Eq. (3), for each these 12 second-nearest neighbor directions; and (iii) 
the associated (averaged) ${\bf m}_i\times {\bf m}_j$ cross product for each of these 12 directions. 
It is important to notice that (1) there two (opposite) directions, i.e. [0$\bar{1}$1] and [0$1\bar{1}$] that have the largest relative spin-current energy 
of the order $w \simeq 0.20$; (2) four other directions -- that are [101],   [$\bar{1}$01],  [$\bar{1}$0$\bar{1}$] and [10$\bar{1}$] -- have significant relative spin-current energies of
approximately {\it half} the predominant one, $w/2 \simeq 0.10$; (3) out of these six directions, three of them 
(namely,  [0$\bar{1}$1],  [101],   [$\bar{1}$01])
have an  averaged ${\bf m}_i\times {\bf m}_j$ close to be ${\bf \zeta} \simeq 4(3\hat{\bf x}-2\hat{\bf y}-\hat{\bf z})$, while the other three have a nearly 
{\it opposite}  averaged ${\bf m}_i\times {\bf m}_j$.
Taking into account items (1)-(3), only considering the six aforementioned directions (as shown in Table I,  the other six second-nearest neighbor directions have smaller relative spin-current energies, and thus can be neglected in first approximation)  
and using commutative properties of the mixed product  allow the rewriting of the last term of Eq. (3) as:
\begin{align}
\Delta E = - C \sum_i {\bf u}_i \cdot \left [ \left ( w {\bf e_{[0\bar{1}1]}}-w{\bf e_{[01\bar{1}]}} + 
\frac{w}{2} {\bf e_{[101]}}-\frac{w}{2} {\bf e_{[\bar{1}0\bar{1}]}}
+\frac{w}{2} {\bf e_{[\bar{1}01]}}-\frac{w}{2} {\bf e_{[10\bar{1}]}} \right)\times {\bf \zeta} \right]
\end{align} 
where ${\bf e_{[0\bar{1}1]}}$, ${\bf e_{[01\bar{1}]}}$,   
${\bf e_{[101]}}$, ${\bf e_{[\bar{1}0\bar{1}]}}$, 
${\bf e_{[\bar{1}01]}}$, ${\bf e_{[10\bar{1}]}}$ are unit vectors
along the $[0\bar{1}1]$, $[01\bar{1}]$,   
$[101]$, $[\bar{1}0\bar{1}]$, 
and $[\bar{1}01]$, $[10\bar{1}]$ directions, respectively.
As a result, Equation (4) becomes:
\begin{align}
\Delta E = - \sqrt2 C w \sum_{i} {\bf u}_i \cdot \left [\left ( -\hat{\bf y}+2\hat{\bf z} \right) 
  \times {\bf \zeta} \right ]
\end{align} 
Once knowing that we also numerically found that, in the type-II cycloid, the cross product  ${\bf m}_i\times {\bf m}_j$ between two adjacent 
$i$ and $j$ sites along the [0$\bar{1}$2] direction 
is also equal to  ${\bf \zeta}$, Equation (5) then naturally explains why, in the type-II cycloid, (i)  the propagation direction of the magnetic cycloid  is now along [0$\bar{1}$2]; 
and (ii) the magnetic moments rotate in the plane defined by (the homogeneous) ${\bf u_i}$ and $-\hat{\bf y}+2\hat{\bf z}$. In other words, this type-II cycloid can be thought as being 
the result of a specific combination (i.e.,  with very specific weights) of {\it different} individual solutions of the spin-current model- each individual solution corresponding to a given second-nearest neighbor direction. In contrast,  the type-I cycloid is associated with a {\it single} individual solution.  We are not aware that such remarkable magnetic organization has ever been proposed or found in the literature. 
This combination occurs for large value of $C$ because the last term of Eq. (3) indicates that the more second-nearest-neighbor directions participate 
in this term the most likely the corresponding energy can be lowered (note that not {\it all} the second nearest neighbor directions can 
equally participate in this last term, since this would lead to an exact cancellation of all the sums involved in the last term of Eq. (3) for an homogeneous polarization).
Interestingly, Eq. (5) also tells us why the polarization is not anymore along the [111] direction when the type-II cycloid forms. 
As a matter of fact, plugging the previously determined ${\bf \zeta} \simeq 4(3\hat{\bf x}-2\hat{\bf y}-\hat{\bf z})$ into this latter equation gives:
\begin{align}
\Delta E = - \sqrt2 C w \sum_{i} {\bf u}_i \cdot \left [5\hat{\bf x}
+6\hat{\bf y}+3\hat{\bf z}  \right ]
\end{align} 
 Such formula can be thought as representating a coupling between the local electric dipoles
 $Z^* {\bf u}_i$ (where $Z*$ is the Born effective charge) and 
 a ``magnetically-induced'' electric field that is equal to ${\bf \cal E}=\frac{\sqrt2 C w}{Z^*} \left [5\hat{\bf x}
+6\hat{\bf y}+3\hat{\bf z}  \right ]$. As indicated by its Cartesian components, this electric field
wants to favor a polarization having a larger y-component, a smaller z-component and an intermediate x-component.  This is consistent with the simulations shown 
in Fig. 2c, and therefore explains why the system becomes triclinic when the type-II cycloid forms. 
It is also very  likely that such phase transition can lead to large physical responses since, e.g., 
giant piezoelectric and dielectric coefficients have been found in 
low-symmetry phases \cite{noheda,pmn-pt,HeffPZT1,HeffPZT2,AaaronPSN,JorgePSN}.

In summary, our calculations reveal that, and explain why, original magnetic arrangements and an unusual, magnetically-induced  phase transition between a high-symmetry and a low-symmetry ferroelectric phase (that is accompanied by a rotation of the electrical polarization) can occur in systems possessing  strong (spin-lattice) spin-current interactions \cite{HeffA}.  Discovering such systems may occur by either (i)  considering multiferroic nanostructures, since, e.g., BFO films under various epitaxial conditions have been recently found to exhibit three different magnetic configurations \cite{Sando}; or (ii) by studying multiferroics made of elements having strong spin-orbit coupling, because this latter
coupling is at the heart of the spin-current model \cite{Katsura1,Aldo,PRL1} (note that transitions from low-spin to high spin-state should also result in an increase of the strength of the  spin-current interaction, or, conversely, reducing the magnitude of the magnetic moments can be thought as reducing the strength of the spin-current interaction, according to Eq. (1)). 
 We therefore hope that the present work will encourage the
discovery of such multiferroics, and deepens the current knowledge of these fascinating materials.

%-----------------------Figure here--------------------------------------------------------------------------------------------------------
%\begin{figure}
%\includegraphics[width=75mm,height=85mm]{../Figs/Fig2.eps}
%\caption{(Color online) Temperature dependence of the three Cartesian components of the supercell average of the local mode for three different values of the $C$ coefficient. The magnetic and paraelectric--to--ferroelectric transition temperatures, $T_N$ and $T_C$, are shown by dashed and solid vertical lines, respectively.
%Panel (a) corresponds to 
%$C=7\times 10^{-6}$ Hartree/Bohr$\mu_B^2$, which leads to the spin-canted magnetic structure below $T_N$. Panel (b) is associated with 
%$C=2.8\times10^{-5}$Hartree/Bohr$\mu_B^2$, which yields the type-I cycloid below $T_N$.
%Panel (c) represents the results for $C=6.3\times10^{-5}$ 
%Hartree/Bohr$\mu_B^2$, whose value generates a type-II cycloid below $T_N$.
%It can be seen that the rhombohedral polar phase survives only for a small window of temperature, before transforming to a triclinic polar state, for the largest shown $C$ value.}
%\label{Fig.2}
%\end{figure}
%------------------------Figure here----------------------------------------------------------------------------------------------------------

\par

We thank ARO Grant No. W911NF-12-1-0085 for personnel support.
Office of Basic Energy Sciences, under contract ER-46612,
ONR Grants No. N00014-11-1-0384 and N00014-12-1-1034, and NSF Grant  
No. DMR-1066158 are also acknowledged for discussions with
scientists sponsored by these grants.

\newpage
%-----------------------Figure here--------------------------------------------------------------------------------------------------------
%-----------------------Figure here--------------------------------------------------------------------------------------------------------
\begin{figure}
\includegraphics[width=135mm,height=125mm]{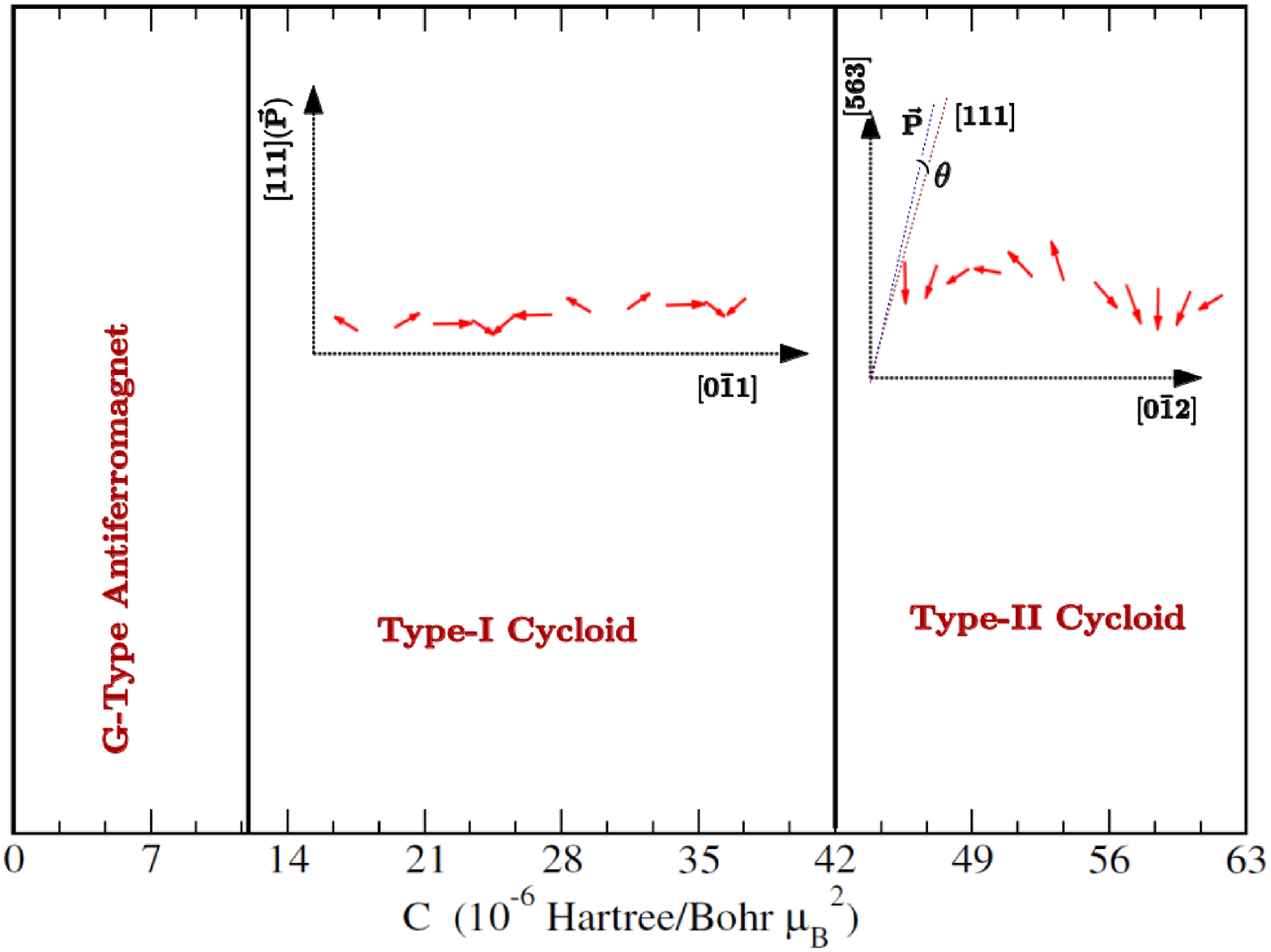}
\caption{(Color online) Range of the different magnetic structures numerically found when varying the spin-lattice coupling constant C. A snapshot of the different cycloids in their cycloidal plane is also provided by means of red arrows (the snapshot corresponds to $C = 1.4\times 10^{-5}$ Hartree/Bohr$\mu_B^2$ for the type-I cycloid and to $C = 5.6\times 10^{-5}$ Hartree/Bohr$\mu_B^2$ for the type-II cycloid). The directions of the cycloidal propagation, of the polarization and of  the  pseudo-cubic [111] axis are also schematized there, in order to emphasize the difference between the two cycloids.}
\label{Fig.1}
\end{figure}
%------------------------Figure here----------------------------------------------------------------------------------------------------------
\newpage
\begin{figure}
\includegraphics[width=135mm,height=125mm]{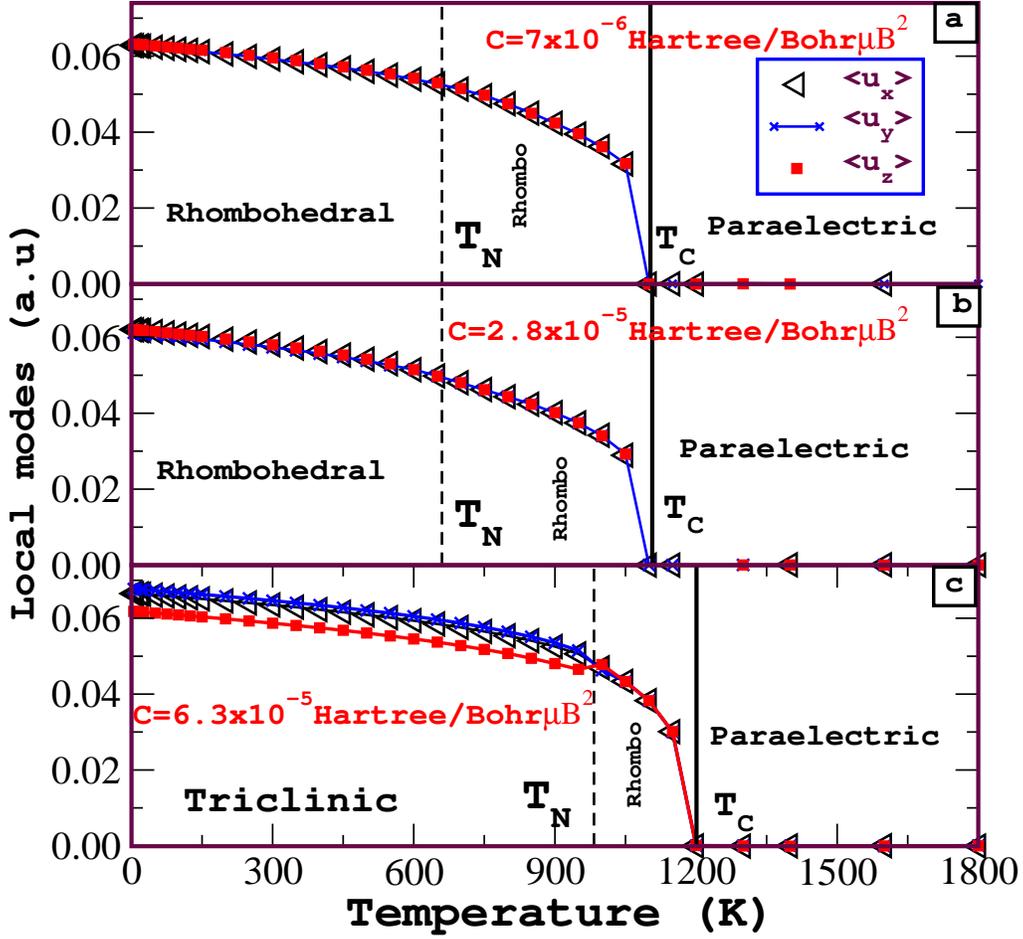}
\caption{(Color online) Temperature dependence of the three Cartesian components of the supercell average of the local mode for three different values of the $C$ coefficient. The magnetic and paraelectric--to--ferroelectric transition temperatures, $T_N$ and $T_C$, are shown by dashed and solid vertical lines, respectively.
Panel (a) corresponds to 
$C=7\times 10^{-6}$ Hartree/Bohr$\mu_B^2$, which leads to the spin-canted magnetic structure below $T_N$. Panel (b) is associated with 
$C=2.8\times10^{-5}$Hartree/Bohr$\mu_B^2$, which yields the type-I cycloid below $T_N$.
Panel (c) represents the results for $C=6.3\times10^{-5}$ 
Hartree/Bohr$\mu_B^2$, whose value generates a type-II cycloid below $T_N$.
It can be seen that the rhombohedral polar phase survives only for a small window of temperature (shown as Rhombo), before transforming to a triclinic polar state, for the largest shown $C$ value.}
\label{Fig.2}
\end{figure}

\begin{thebibliography}{99}
\bibitem{DM} Note that the spin-current model can also be seen as a Dzyaloshinsky-Moriya (DM) interaction \cite{Dzyaloshinsky,Moriya}, which is given by ${\bf D'}  \cdot({\bf m}_i\times {\bf m}_j)$, with the DM vector, ${\bf D'}$, being equal to  $-C ( {\bf P} \times {\bf e_{ij}})$.
\bibitem{footnoteSDW} Note that there is also a spin density wave, involving small out-of-plane components of the magnetic moments, 
that occurs in BFO bulks, in addition to the in-plane cycloid \cite{SDW,PRL1}.
\end{thebibliography}
\end{document}